\documentstyle[12pt]{article}
\begin{document}
\thispagestyle{empty}
\baselineskip=15pt

\begin{center}
{\bf The Critical Scale of the Dual Superconductor Picture of QCD}
\end{center}
\begin{center}
\underline{H. Ichie} and  H. Suganuma\\
\vspace{0.7cm}
{\it 
Research Center for Nuclear Physics(RCNP),\\
Osaka University,Ibaraki,567 Japan}\\

\end{center}

\begin{quote}
{\small
   The nonperturbative phenomena of QCD like color confinement
is well described  through the dual superconductor picture in the Maximally 
Abelian (MA) gauge. 
In this gauge,
monopoles appear as important degrees of freedom composed by
nonabelian gauge fields.
We investigate the peculiar size $R_c$ of the monopole in the MA gauge,
considering the theoretical similarity between spinglass and the  MA gauge 
fixing.
As for the properties at larger scale than the monopole size $R_c$,
the system can be described by the dual superconductor theory  with  local
abelian fields.
At  shorter distances than $R_c$,
such a dual Higgs theory should be treated as a non-local field theory,
and therefore one must take another framework like the 
perturbative QCD. 
Thus, the monopole size gives a critical scale on the change of the 
theoretical structure of QCD.

}
\end{quote}
\section{Introduction}

\indent\indent The strong interaction is subjected to  Quantum 
Chromodynamics (QCD).
Due to the nonabelian gauge theory, the gauge coupling becomes very strong in 
 the low-energy region, while it is weak in the high-energy region.
Accordingly, QCD phenomena are divided into two theoretical categories:
the strong coupling leads to  complicated nonperturbative phenomena such as
color confinement and chiral symmetry breaking, while
high-energy phenomena are understood by the perturbative QCD.

Since there is no well-established analytical method for
nonperturbative phenomena,
one must carry out the Monte Carlo simulation based on  the lattice QCD
or apply the effective models described by the relevant degrees of freedom 
for the low-energy physics. 
As for the chiral dynamics, 
the pion and the sigma meson, which are bound-state of quarks, play an important role
for an infrared effective theory  such as 
Nambu-Jona-Lasinio model and (non-)linear sigma model.
On the other hand, confinement is essentially described by the dynamics of 
gluons rather than quarks.

In 1981, 't Hooft proposed an interesting idea that QCD  
 is reduced to an 
abelian gauge theory including monopoles in the abelian gauge \cite{thooft}, and confinement is 
realized by monopole condensation:
the electric field is excluded from the QCD vacuum by the dual Meissner effect,
which is regarded as a dual version of the superconducting 
theory \cite{suzuki2,suganuma1,ichie}.
In the abelian gauge, only abelian variables (abelian gluons and monopoles)
are essential for confinement \cite{suzuki1,hioki},
while the off-diagonal gluons seem irrelevant except at the short distance.
The nonabelian nature appears as the existence of 
monopoles in the low-energy region.
Therefore, the relevant degrees of freedom are changed from  SU(N$_c$)
gluons into abelian gluons and monopoles in this framework.

In principal, these infrared variables like monopoles are described as 
composite operators in terms of original variables (gluons) in QCD so that
the infrared variables are considered to have their 
peculiar size $R_c$ like hadrons.
At the large scale where $R_c$ can be neglected, these infrared variables 
can be treated as {\it  local} fields, and the system can be described by 
a local field theory \cite{ezawa}.
On the other hand, at shorter distances than $R_c$, the {\it 
nonlocality} due to the size effect appears explicitly, and the effective 
theory should not be workable  as a local field theory.

In this paper, we study the critical scale of the dual superconductor theory
of QCD in terms of the {\it nonlocality} originating from the peculiar size
 of the infrared variables in the  Maximally Abelian (MA) gauge using 
 the lattice QCD.  
For simplicity, we concentrate ourselves on the $N_c=2$ case hereafter. 

\section{Monopoles as the infrared variables}

\indent\indent Recent lattice QCD simulations show that the confinement phenomena can be
described by monopole condensation  in the MA gauge \cite{kronfeld}.
In terms of the link variable $U_\mu(s) \equiv U_\mu^0(s) + i\tau^a 
U_\mu^a(s)$, the MA gauge fixing is defined by maximizing 
\begin{eqnarray}
R \equiv - H \equiv \sum_{s,\mu} {\rm tr}( U_\mu(s) \tau_3 U^{\dagger}_\mu(s) \tau_3 ) 
= \sum_{s,\mu}\{ 
(U^0_\mu(s))^2+(U^3_\mu(s))^2-(U^1_\mu(s))^2-(U^2_\mu(s))^2 \}
\end{eqnarray}
through the gauge transformation.
Here, $H$  corresponds to the hamiltonian of spinglass, which is 
discussed in Sec.3.  
In this gauge, the off-diagonal components, $U_\mu^1$ and $U_\mu^2$, are 
forced to be small,
and therefore the QCD system seems describable by U(1)-like variables 
approximately.
It is to be noted that the MA gauge is a sort of the abelian gauge, because
$X(s) \equiv \sum _{\mu,\pm } U_{\pm \mu}(s) \tau_3 U^{\dagger}_{\pm \mu}(s)$
is diagonalized there. (For the simple notation, we use $U_{-\mu}(s) \equiv 
U^{\dagger}_\mu(s-\mu)$ in this paper).

First, we extract the U(1)-variable from $U_\mu(s) \in $SU(2) and define 
the monopole current in the MA gauge. The SU(2) link variable $U_\mu$(s)
is factorized as 
\begin{eqnarray}
U_\mu(s) = 
\left( 
\begin{array}{cc} \sqrt{1-|c_\mu(s)|^2}  & - c^{*}_\mu(s)
 \\
 c_\mu(s)  & \sqrt{1-|c_\mu(s)|^2} \end{array}
\right) 
\left( \begin{array}{cc} e^{i\theta_\mu(s)} & 0 \\ 0 & e^{-i\theta_\mu(s)}
 \end{array} \right) 
= M_\mu (s) u_\mu(s),
\end{eqnarray}
where $u_\mu(s)$ and  $M_\mu(s)$ correspond to the diagonal part and the 
off-diagonal part, respectively.
For the residual U(1)$_3$ gauge transformation, $u_\mu(s)$ behaves as the 
abelian gauge field, while $c_\mu(s)$ behaves as the charged matter:
\begin{eqnarray}  
u_\mu(s) & \rightarrow & \omega(s) u_\mu(s) \omega^{\dagger}(s+\mu), 
\nonumber \\
c_\mu(s) & \rightarrow & {\rm e}^{ -i\alpha (s) }  c_\mu(s),
\end{eqnarray}
where $\omega (s) \equiv e^{i \alpha(s) \tau_3/2}$ $\in$ U(1)$_3$ is the gauge 
function.
In the MA gauge, the nonperturbative phenomena can be described well by 
the U(1)-link variable $u_\mu(s)$ alone, which is called as {\it Abelian 
Dominance} \cite{suzuki1}.

Taking the angle $\theta _\mu(s)$ as $-\pi \le \theta_\mu(s) < \pi$,
the 2-form $(\partial \wedge \theta)_{\mu\nu}$ is divided into two 
parts \cite{degrand},
\begin{eqnarray}
(\partial \wedge \theta )_{\mu\nu} = \bar \theta_{\mu\nu} + 2 \pi n_{\mu\nu},
\end{eqnarray}
where $n_{\mu\nu} \in Z$ and $\bar \theta_{\mu\nu}$ is defined as
\begin{eqnarray}
\bar \theta_{\mu\nu}\equiv {\rm mod}_{2 \pi} (\partial \wedge \theta)_{\mu\nu} \in 
[-\pi,\pi).
\end{eqnarray}
As for  the relation to the continuum field variables,
the lattice variables are written as $\theta_\mu = a e A_\mu /2$,
$\bar \theta_{\mu\nu} = a^2 e f_{\mu\nu} /2$ and 
$ 2 \pi n_{\mu\nu} = a^2 e f_{\mu\nu}^{\rm sing}/2$
with lattice spacing $a$.
Here, $A_\mu$ is the abelian gauge field; $f_{\mu\nu}$ and 
$f_{\mu\nu}^{\rm sing}$ are the regular and singular parts of the field 
strength tensor, respectively.
In the continuum limit $a \rightarrow 0$, 
$f_{\mu\nu}$ takes a finite value, while $f^{\rm sing} = 
4\pi n_{\mu\nu}/(ea^2)$ goes to infinity and corresponds to 
the Dirac string \cite{degrand}.

As a remarkable fact, there appear monopoles in the MA gauge reflecting 
the nonabelian nature in QCD \cite{kronfeld}.
Here, the monopole originates from the nontrivial gauge transformation 
corresponding to  $\Pi_2(\mbox{SU(2)/U(1)})=Z_\infty$.
The monopole current is defined on the lattice as 
\begin{eqnarray}
k_\mu \equiv \frac{1}{2\pi} \partial^\nu {}^* \bar \theta_{\nu\mu} 
= - \partial^\nu  {}^{*}n_{\nu\mu}.
\end{eqnarray}
As for the relevant role of the monopole,
lattice QCD simulations  indicate that nonperturbative phenomena 
like confinement and chiral symmetry breaking are brought by the contribution 
of the monopole in the MA gauge, which is called as  {\it Monopole 
Dominance}
\cite{miyamura}.

In the MA gauge, there are several advanced points on the extraction of
the U(1)-variables.
The SU(2) link variable $U_\mu(s)$ can be approximated by U(1)-link variable
$u_\mu(s)$ in the MA gauge, $U_\mu \simeq u_\mu$, because of the 
reduction of the  off-diagonal components.
Therefore, the U(1) action $\sum_{s\mu\nu}(1-\frac12{\rm tr} e^{i\bar 
\theta_{\mu\nu}(s) \tau_3} )$ takes a small value like the SU(2) action.
In the continuum limit $a \rightarrow 0$, $\bar \theta_{\mu\nu}$ becomes 
small so that ultraviolet fluctuation is suppressed in the U(1) field 
strength $f_{\mu\nu}= 2 \bar \theta_{\mu\nu}/(e a^2)$.
Thus, the total field strength $(\partial \wedge \theta)_{\mu\nu}$ is
almost descritized as $2 \pi n_{\mu\nu}$, and hence the Dirac-string 
contribution is clearly extracted at each praquette.
Accordingly, the monopole current is definitely obtained on the lattice 
in the MA gauge. The confinement phenomena are well described by 
the monopole current in the MA gauge \cite{kronfeld}.

\section{Non-locality in  the MA Gauge}

\indent\indent In this section, we investigate the procedure of the MA gauge fixing more detail to 
understand how  the nonlocality appears in the infrared variables, e.g. 
$u_\mu(s)$ and  $k_\mu(s)$.
The functional $R[U_\mu(s)]$, which is to be maximized in the MA gauge,
is transformed by the gauge function $V(s)$ as 
\begin{eqnarray}
R \rightarrow R & = &  \sum_{\mu\pm}
{\rm tr} \{ V(s) U_\mu(s) V^{\dagger}(s \pm \mu) \tau_3 V(s \pm \mu) U^{\dagger}_\mu(s)
V^{\dagger}(s) \tau_3 \} \nonumber \\
& = & \sum_{\mu\pm}
{\rm tr} \{  U_\mu(s) \phi(s \pm \mu)  U^{\dagger}_\mu(s) \phi(s) \} \nonumber \\
& = &  \sum_{\mu\pm}
\phi^a(s) g^{ab}_\mu(s) \phi^b(s \pm \mu),  
\end{eqnarray}
with
\begin{eqnarray}
g_\mu^{ab}(s) = \{ (U^0_\mu(s))^2 -( U^i_\mu(s))^2) \}\delta^{ab}
+  U^a_\mu(s) \cdot  U^b_\mu(s). 
\end{eqnarray}
Here, we define the `spin' variable $\phi(s) \equiv
\phi^a \tau^a \equiv V^{\dagger}(s) \tau_3 V(s) $ is determined by 
maximizing $R$, and plays a similar role to the Higgs field in the 't Hooft-Polyakov monopole.
This system can be regard as the classical `spin' system obeying the 
Hamiltonian $H=\sum_{s\mu} g(s) {\bf s}_\mu \cdot {\bf s}_{s+\mu}$,
which should be minimized in the ground state. 
 The most remarkable feature of this system is that
the spin interaction are random depending on the links, 
which is  known to be the spinglass.
Due to the randomness of the  interaction, there are many local minima of
 almost the same `energy' in the configuration space.
Therefore, even the small fluctuation of the interaction has
the influence with surrounding `spins' in the large region.
Such  instability leads to the macroscopic correlation.
Thus, the nonlocality in $U^{\rm MA}_\mu(s)$ appears in the MA gauge.

\section{Numerical Calculation}

\indent\indent In general, \{$U_\mu^{\rm MA}$\} is expressed as a function of
\{$U_\mu$\}, however, the correspondence between $U_\mu^{\rm MA}$ and $U_\mu$
is not local relation.
On the contrary, $U_\mu^{\rm MA}(s_0)$ on a link is composed by many 
original link variables $U_\mu(s)$ with $s$ belonging an extended region 
around $s_0$.
Such an extension leads to the nonlocality of the infrared variables
in the MA gauge.
In this paper, we investigate extension numerically  using lattice
simulation.

We examine the influence of the local change of $U_{\mu_{\small 0}}(s_0)$ to 
$U_{\mu}^{\rm MA}(s)$ around $s_0$.
We first prepare two link configurations \{$U_\mu(s)$\} and \{ 
$U^{'}_\mu(s)$\} before 
the MA gauge fixing:  \{$U^{'}_\mu(s)$\} is defined
by changing $U_\mu(s)$ at a certain link $s_0$,
\begin{eqnarray}
 U^{'}_\mu(s)=U_\mu(s) + \delta \tilde U_\mu(s) 
\hspace{2cm} \mbox{at $(s,\mu) =(s_0,\mu_0)$}  \\
 U^{'}_\mu(s)=U_\mu(s) \hspace{3.7cm} \mbox{at $(s,\mu) \ne (s_0,\mu_0)$}. 
\end{eqnarray}
Second, we perform the MA gauge fixing for these two configurations 
respectively,
and obtain \{$U^{\rm MA}_\mu(s)$\} and \{$U^{'\rm MA}_\mu(s)$\}.
We then estimate the `distance' between $U^{\rm MA}_\mu(s)$ and
 $U^{'\rm MA}_\mu(s)$,
\begin{eqnarray}
d_{U}(s,\mu;s_0,\mu_0) \equiv 1- \frac12 {\rm tr}[U_\mu^{\rm MA}(s)
U_\mu^{'\dagger {\rm MA}}(s)].
\end{eqnarray}
This indicate the difference between $U^{\rm MA}_\mu(s)$ and
 $U^{'\rm MA}_\mu(s)$, and $d_U$ goes to zero as $U^{\rm MA}_\mu(s) = 
 U^{'\rm MA}_\mu(s)$.
As the residual gauge-invariant variable, we estimate also the  
`distance' between $\bar \theta^{\rm MA}$ and $\bar \theta^{'\rm MA}$,
\begin{eqnarray}
d_{\bar \theta} (s,s_0) \equiv \sum_{\nu \ne \mu_0}
d_{\bar \theta} (s,\mu,\nu;s_0,\mu_0), 
\end{eqnarray}
where
\begin{eqnarray}
d_{\bar \theta} (s,\mu,\nu;s_0,\mu_0) \equiv  1 - \frac12 
{\rm tr[exp}(i |\bar \theta^{\rm MA}_{\mu\nu}(s)|-i|\bar \theta'^{\rm MA}_
{\mu\nu}(s)|)].
\end{eqnarray}

Numerical simulation has been performed on $16^4$ lattice with $\beta=2.4$
using 433  samples from  9 gauge configurations.
The numerical results of $d_U$ and $d_{\bar \theta}$  are shown in 
Fig.1 (a) and (b), respectively.
The data can be fitted by exponatial curves denoted by dotted lines.
From these results, the variable $U^{\rm MA}_\mu(s)$ 
in the MA gauge 
would be expressed as the nonlocal function of $U_\mu(s)$ in the extended 
region with the size of $R_c$ around $s_0$.
The variable $U_\mu^{\rm MA}(s_0)$ at a point $s_0$ is constructed
by $U_\mu(s)$ in the nonlocal region around $s_0$ with radius $R_c$.
Thus, $U_\mu^{\rm MA}(s)$ has a peculiar size of $R_c$ in terms of original 
link variables.

\section{Concluding  Remarks} 

\indent\indent We  have investigated the nonlocal nature of infrared
variables (abelian variables and monopoles) in the MA gauge.
We first consider  origin of the nonlocality, considering the similarity 
between the MA gauge fixing process and  the theoretical structure of spinglass.
Using the lattice QCD simulations on $16^4$ lattice with $\beta=$2.4, 
 the nonlocal extension of infrared variables in the MA gauge is found to be about 
$R_c \simeq 0.24$fm as terms of the original link variables.
Therefore, at  large distance scale where $R_c$ can be neglected, these 
variables can be treated as local fields, and the
dual superconductor theory would be workable, while  the perturbative QCD
is useful instead at short distance scale.
Thus, the theoretical structure of QCD would be changed around $R_c$.

The simulations are performed by VPP500 of RIKEN.

Fig.1\\
(a) The distance $d_U$  between $U^{\rm MA}_\mu$ and $U^{'\rm MA}_\mu$;
(b) The distance  $d_{\bar \theta}$ between  $\bar \theta^{\rm MA}$ and $\bar \theta^{'\rm MA}$.
The fluctuation given at ${\bf x}={\bf 0}$ has influence with 
surrounding link variables. 
The data can be fitted by exponatial curves denoted by dotted lines.
The extension of infrared variables is about $R_c=1.5a = 0.24$fm with lattice 
constant $a=0.16$fm. 
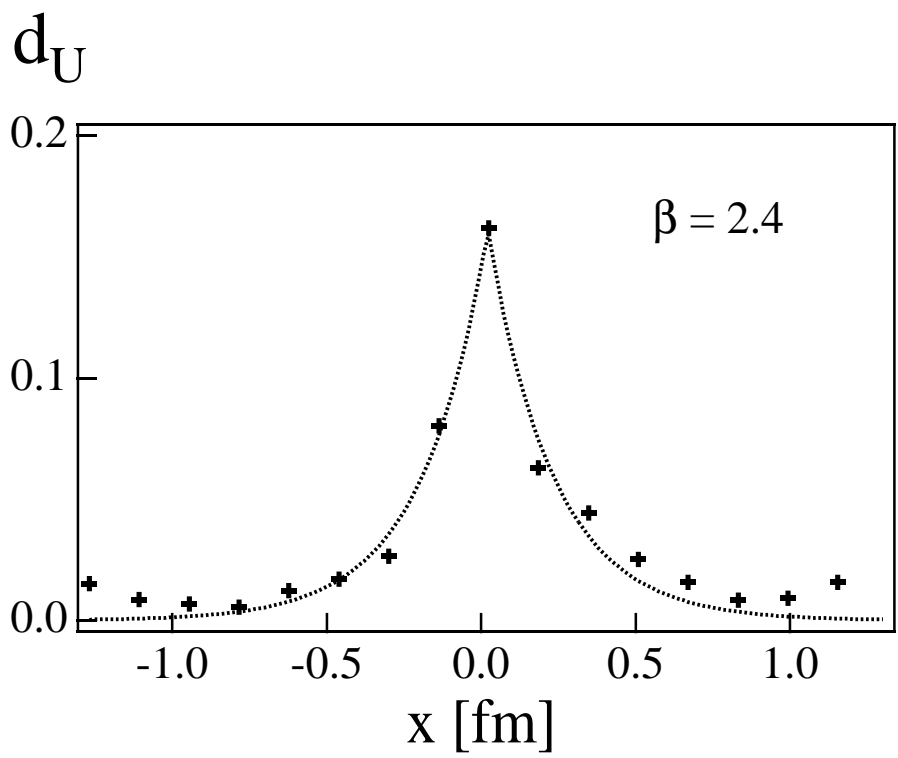
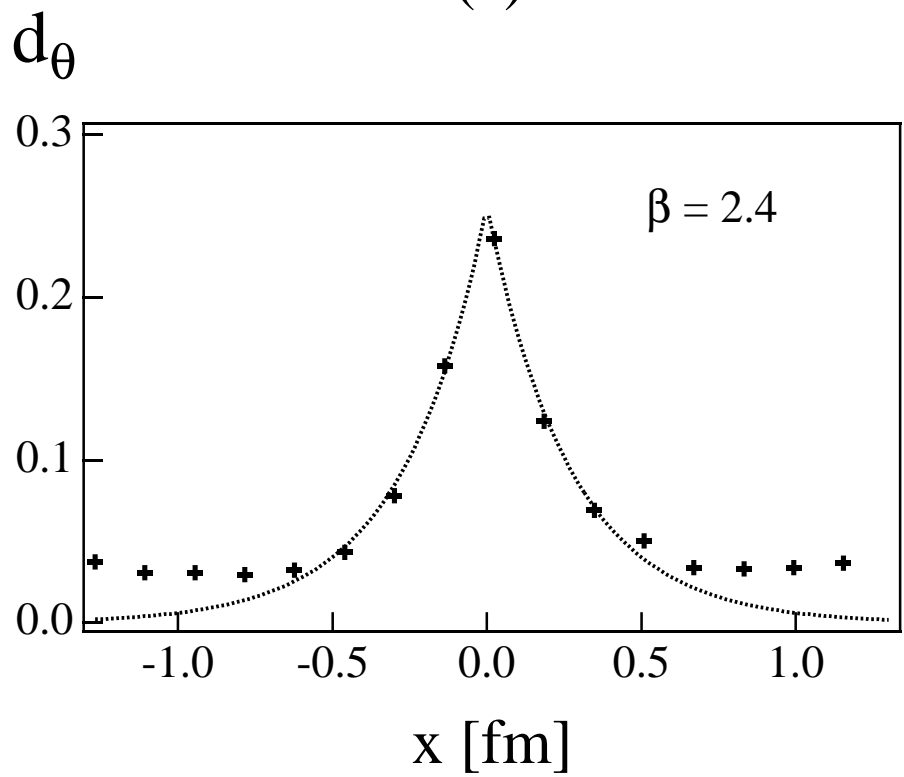
\end{document}